\documentstyle[prl,aps]{revtex}
\begin{document}
\title{Coarse graining and a new strategy for
renormalization} \draft
\author{Ji-Feng Yang}
\address{Department of Physics, East China Normal University,
Shanghai 200062, P. R. China}
\maketitle

\begin{abstract}
We present the natural arguments for the rationality of a recently
proposed simple approach for renormalization which is based
solving differential equations. The renormalization group equation
is also derived in a natural way and recognized as a decoupling
theorem of the UV modes that underlie a QFT. This new strategy has
direct implications to the scheme dependence problem.
\end{abstract}
\section{Introduction}Recently, we proposed a new approach for
calculating radiative corrections without introducing any form of
regulator and any form of removal of UV
divergence\cite{YYY,PRD65,CTP38}, which is a differential equation
approach with ambiguities to be fixed through rational boundary
conditions. In this simple approach, many complicated aspects
associated with conventional regularizations (e.g., the subtle
definition of Dirac matrices $\{\gamma^5, \gamma^{\mu}\}$ and
metric tensor $g^{\mu\nu}$ in dimensional regularization; the
notorious power law divergences in cutoff regularization, and so
on) simply do not show up\cite{CTP38}. It is especially efficient
in the nonperturbative contexts where conventional regularization
and/or subtraction schemes often make it very hard to extract
physical information from the calculated quantities, as the
proposed approach can dramatically reduce the difficulty in
extracting physical information\cite{LOU}. It is also applied in
Ref.\cite{PRD65} to massless $\lambda\phi^4$ to discuss the
problem of nontrivial symmetry breaking solution in various
regularization and renormalization prescriptions.

In this short report, we shall: (1) to present the natural
rationality of the simple approach in Sec. II; (2) sketch a simple
derivation of the renormalization group equation as a natural
decoupling theorem of the underlying short distance modes in Sec.
III. The final section is devoted to discussion and summary. Part
of these arguments have been available in the e-print
form\cite{YYY}.
\section{Underlying theory and finiteness of QFT}
Our starting point is the well known point of view that the
conventional QFT should be replaced by a complete quantum theory
of everything (QTOE) with correct high energy details. The low
energy physics are defined by the coarse grained low energy
sectors of QTOE with the extremely short distance processes
integrated out. The high energy modes' contributions are
physically suppressed by certain physical mechanism defined in
QTOE (unknown to us) rather than 'cut off' by hand. This
understanding naturally motivates the presence of a set of
parameters(denoted as $\left\{ \sigma \right\} $) to characterize
the high energy modes' contributions in the coarse grained
objects. Technically, it is these constants and the way they
appear in the generating functional that suppress the high energy
modes while keep the 'effective' quanta dominant. For this coarse
graining or emergence scenario to be effective, the magnitude of
the parameters in energy unit must be {\bf such that} $\sup
\left\{ \Lambda _{QFT} \right\} \ll \inf \left\{ \sigma \right\} $
with $\Lambda _{QFT}$ representing a general dimensional parameter
(momenta or masses) in the QFT in under consideration.

The preceding magnitude order analysis automatically activates a
limit operation with respect to $\left\{ \sigma \right\} $ on the
coarse grained amplitudes for describing 'low' energy processes,
which will be denoted as $L_{\left\{ \sigma \right\} }$($ \equiv
\lim_{\left\{ \sigma \right\} \rightarrow 0}$ in length unit).
Then the coarse grained vacuum functional in the presence of the
external sources for low energy processes reads \FL
\begin{eqnarray}
&&Z\left( J\left( x\right) |\{\bar{c}\}\right) \equiv L_{\left\{
\sigma \right\} }Z\left( J\left( x\right) |\{\sigma
\}\right)\equiv L_{\left\{ \sigma \right\} }\int D\Phi \left(
x{\bf |}\left\{ \sigma \right\} \right)
\exp \left[ \frac i\hbar S\left( \Phi \left( x%
{\bf |}\left\{ \sigma \right\} \right) ;\left\{ \sigma \right\}
\Vert J\right) \right] ,
\end{eqnarray}
where the $\left\{ \sigma \right\} $ dependence of a function(al)
indicates that they are coarse grained objects well defined in
QTOE. The appearance of the constants $\{\bar{c}\}$ (including
$\bar{\mu}$) in the RHS of Eq.(1) implies that the order of
functional integration and $L_{\left\{ \sigma \right\} }$ can not
be trivially exchanged, otherwise we would get the ill defined
QFT's or divergences, i.e.,
\begin{eqnarray}
L_{\left\{ \sigma \right\} }\int D\Phi \left( x{\bf |}\left\{
\sigma \right\} \right) \exp \left[ \frac i\hbar S\left( \Phi
\left( x{\bf |}\left\{ \sigma \right\} \right) ;\left\{ \sigma
\right\} \Vert J\right) \right] \neq \int D\Phi \left( x\right)
\exp \left[ \frac i\hbar S\left( \Phi \left( x\right) \Vert
J\right) \right] ,
\end{eqnarray}
with $ S\left( \Phi \left( x\right) \Vert J\right) \equiv
L_{\left\{ \sigma \right\} }S\left( \Phi \left( x{\bf |}\left\{
\sigma \right\} \right) ;\left\{ \sigma \right\} \Vert J\right) $
and $ \Phi \left( x\right)\equiv L_{\left\{\sigma
\right\}}\Phi\left( x{\bf |} \left\{ \sigma \right\} \right) $.

In terms of Feynman diagram algorithm, this is (for a one loop
divergent diagram in QFT),\FL
\begin{eqnarray}
L_{\left\{ \sigma \right\} }\Gamma \left( \left( p\right) ,\left(
m\right) ;\{\sigma\}\right) \equiv L_{\left\{ \sigma \right\}
}\int d^DQ\bar{f}_\Gamma \left( Q,\left( p\right) ,\left( m\right)
;\{\sigma \}\right) \neq \int d^DQf_\Gamma \left( Q,\left(
p\right) ,\left( m\right) \right),
\end{eqnarray}
with $f_\Gamma (Q,\left( p\right) ,\left( m\right) )$ being the
integrand of this diagram defined in conventional QFT. The loop
momentum, external momenta and masses are denoted respectively by
$Q,\left( p\right) $and $\left( m\right)$.

In principle we could not evaluate the generating functional or
the Feynman amplitudes without knowing the exact dependence upon
$\{\sigma \}$. However, we can determine each one loop amplitude
(ill defined in QFT) $ L_{\left\{ \sigma \right\} }\Gamma \left(
\left( p\right) ,\left( m\right) ;\{\sigma \}\right) $ up to an
appropriate polynomial of momenta and masses with finite but
undetermined coefficients {\bf as long as} we accept that the QTOE
version of the loop diagram exists.

{\bf THEOREM}. \emph{A one loop amplitude $\Gamma $ defined in
QTOE (ill defined in the conventional QFTs) satisfies the
following kind of natural differential equation, \FL
\begin{eqnarray}
\label{diffeq} \left( {\partial }_p\right) ^{\omega _\Gamma
+1}L_{\left\{ \sigma \right\} } \bar{\Gamma}\left( \left( p\right)
,\left( m\right) |\left\{ \sigma \right\} \right) =\int d^DQ\left(
{\partial }_p\right) ^{\omega _\Gamma +1}f_\Gamma (Q,\left(
p\right) ,\left( m\right) )\ \ \left(\equiv \Gamma ^{\left( \omega
_\Gamma \right) }\left( ( p) ,( m) \right)\right)
\end{eqnarray}
with $\omega _\Gamma $ being the superficial divergence degree or
scaling dimension of such a diagram}.

${\sl Proof}$: Since QTOE is completely well defined, then in any
dimension $D$ of spacetime we have\cite{YYY} $$ \begin{array}{l}
 \left( {\partial }_p\right)^{\omega _\Gamma +1}
 L_{\left\{ \sigma \right\}}\bar{\Gamma}\left(
 \left( p\right) , \left( m\right) |\left\{\sigma \right\}
 \right)  =L_{\left\{ \sigma \right\}}\int d^DQ
 \left( {\partial }_p\right) ^{\omega _\Gamma
+1}\bar{f}_{\bar{\Gamma}}(Q,\left( p\right) , \left(
m\right)|\left\{ \sigma \right\} )\\ =\int
d^DQ\left({\partial}_p\right) ^{\omega _\Gamma +1}L_{\left\{
\sigma \right\} }\bar{f}_{\bar{\Gamma} }(Q,\left( p\right) ,\left(
p\right) |\left\{ \sigma \right\} ) =\int d^DQ\left( {\partial
}_p\right) ^{\omega _\Gamma +1}f_\Gamma (Q,\left( p\right) ,\left(
m\right) )\ \ \ \ Q.E.D.
\end{array}$$
Similar differential equations also hold with ${\partial }_p$
replaced by $\partial _m$. The key observation here is that
differentiating a Feynman amplitude with respect to external
parameters lower the divergence degrees of the
amplitude\cite{dif}.

The solutions to such differential equations are easy to obtain as
\FL
\begin{eqnarray}
\label{solution} &&\Gamma \left( ( p) ,( m);\{\bar{c}\}
\right)\equiv L_{\left\{ \sigma \right\} }\bar{\Gamma}\left(
\left( p\right) ,\left( m\right) |\left\{ \sigma \right\}
\right)\doteq \left( \int_p\right) ^{\omega _\Gamma +1} \Gamma
^{\left( \omega _\Gamma \right) }\left( ( p) ,( m)
\right)\nonumber \\
&&=\left( \int_p\right) ^{\omega _\Gamma +1}\int d^DQ\left(
{\partial }_p\right) ^{\omega _\Gamma +1}f_\Gamma (Q,\left(
p\right) ,\left( m\right) )
\end{eqnarray}
with the symbol '$\doteq $' indicating that the two sides are
equal up to certain integration constants in a polynomial of
momenta and masses of power $\omega _\Gamma $. To determine the
integration constants (which is definitely defined as
$\{\bar{c}\}$ in QTOE from the limit operation$L_{\left\{ \sigma
\right\} }$) we need 'boundary conditions' like symmetries, sum
rules and finally experimental data, which parallels the procedure
of choosing renormalization conditions. For later convenience we
note that there must be a dimensional constant characterizing the
typical length or energy of the QFT under consideration and we
denote it as $\bar{\mu}$. Eq.(~\ref{diffeq}) or (~\ref{solution})
is just our general recipe for evaluating the Feynman amplitudes
that dispenses the notorious divergences and the associated
subtraction. This recipe works in the same way for multiloop
diagrams, for details please refer to Ref.\cite {YYY}. The
guideline is to insert a pair of $\left( \int_p\right) ^{\omega
_\Gamma +1}$ and $\left( {\partial }_p\right) ^{\omega _\Gamma
+1}$ to the two sides of each divergent loop integration as
$L_{\left\{ \sigma \right\} } $ crosses the loop integration from
the left until the $L_{\left\{ \sigma \right\} }$ is finally
removed from all loops in the diagram. For convergent loops
$L_{\left\{ \sigma \right\} }$ can safely cross the loop
integrations. However, by defining that $\left( \partial \right)
^n\equiv {\left( \int \right) }^{ |n|},\ {\left( \int \right)
}^n\equiv \left( \partial \right) ^{|n|},\ $for$\ n<0,{\left( \int
\right) }^n=\left( \partial \right) ^n=1,\ $ for$\ n=0$ and noting
that $\left( \int \right) ^n\times \left(
\partial \right) ^{n}=\left( \partial \right) ^{|n|}\times \left(
\int \right) ^{|n|}=1 $ for $n<0$ we can also put a convergent
loop into the form of Eq.(5) with now $\omega _\Gamma $ denoting
the negative scale dimension of the convergent loop diagram.

We emphasize that the above expressions are correct {\bf provided}
the magnitude order $\sup \left\{ |p|,m,\bar{\mu}\right\} \ll \inf
\left\{ \sigma \right\} $ is satisfied, no matter how large the
mass or momentum is. It is clear that no subtraction is necessary,
no infinite counterterms and bare parameters is present except
finite 'bare' parameters---the tree parameters in Lagrangian. It
is also evident that our strategy is obviously applicable to any
interactions (fields with any spin) in any spacetime, even for
nonlocal interactions, as our deduction does not need any
specifics about interaction. Even the Lorentz invariance and other
symmetry status are not needed at all, as long as the whole
dynamics are consistently defined.

Among the integration constants (which will be denoted as $\left\{
C\right\} $ in contrast to $ \left\{ \bar{c}\right\} $), there
must be a dimensional scale to balance the dimensions in the
logarithmic function of momenta (which will be denoted as $ \mu
_{int}$ that corresponds to $\bar{\mu}$). The integration
constants $ \left\{ C\right\} $ span a space in which the QTOE
prediction $\left\{ \bar{c }\right\} $ just lies on one point of
this space. Obviously, the QTOE definition of the Lagrangian
constants and the 'loop' constants $\left\{ \bar{c}\right\} $
(including $\bar{\mu}$) should be scheme and scale
invariant\cite{scheme,Maxw}. This may accentuate and accelerate
the extraction of physical parameters out of renormalization
scheme and scale dependent parametrization\cite{Maxw}, namely,
once we fixed the Lagrangian parameters in some physical way, we
can in principle systematically extract $\left\{ \bar{c}\right\} $
from experiments. Thus inequivalent choices of $\left\{ C\right\}
$ would correspond to different physics. (Note that here the words
'bare parameters' does not mean no interaction. The 'bare' or tree
parameters in QFT in fact characterize the 'elementary' quantum
dynamics of the low energy processes in the lagrangian level.)

For perturbative Feynman diagram representation, our differential
equation approach is similar to the celebrated BPHZ
algorithm\cite{BPHZ}. However, we must point out that: 1)in
practice one must first specify a regularization scheme (that is,
introducing certain kind of artificiality in the computation)
before BPHZ is implemented; 2)the subtraction procedure in BPHZ
could only lead to special set of constants that solve the
differential equations for relevant Feynman amplitudes; 3)the BPHZ
program ends up with the introduction of infinite bare quantities
while there is no room for such infinite quantities at all if one
adopts the underlying theory standpoint; 4) the application of
BPHZ (and other conventional programs) in nonperturbative
circumstances is rather involved that might preclude any useful
(or trustworthy) predictions, while the differential equation
approach makes the calculation easier and the physical predictions
more accessible\cite{LOU}. We think our differential equation
approach (and the underlying theory scenario) generalizes, refines
or improves various conventional renormalization programs in a
natural way. Moreover, we could get rid of the various
shortcomings in conventional programs {\em due to} inevitable
introduction of a regularization scheme, an artificial substitute
for the true short distance physics. This drawback is especially
troublesome in nonperturbative applications\cite{KSW}.
\section{Renormalization group
equation and decoupling of underlying modes} From the preceding
discussions on the constants $\{\bar{c}\}$, we can parametrize
them in such a way that $\{\bar{c}\}=\{\bar{\mu},\left[ \bar{c}
^0\right] \},$ dim$\left\{ \bar{c}^0\right\} =0,\partial
_{\bar{\mu}}\bar{c} ^0=\partial _g\bar{c}^0=0,\forall
\bar{c}^0,\forall g:$ dim$\left\{ g\right\} \neq 0$,i.e., we
parametrize all the dimensional constants in $\{\bar{c}\}$ as
$\{\bar{\mu}{\bar{c}}^0\}$ with $\{\bar{c}^0\}$ dimensionless
constants. This is legitimate as $\{\bar{c}\}$ are all of the same
order as $\{g\}$, otherwise they should belong to the same set of
the underlying constants $\{\sigma\}$ and vanish from the explicit
formulation of QFTs.

Rescaling every dimensional parameters in a general vertex
function $\Gamma ^{\left( n\right) }\left( \left( p\right) ,\left(
g\right) ;\{\bar{c}\}\right) $ (we denote masses and couplings
collectively as $ \left( g\right) $) that is well defined in QTOE,
we have \FL
\begin{equation}
\label{CSE} \left\{ s\partial _s+\Sigma d_gg\partial
_g+\bar{\mu}\partial _{\bar{\mu} }-d_{\Gamma ^{(n)}}\right\}
\Gamma ^{(n)}(\left( sp\right) ,\left( g\right) ;\{\bar{c}\})=0.
\end{equation}
with $d_{\cdots }$ denoting the mass dimensions of the associated
constants. Since all the constants $\{\bar{c}\}$ only appear in
the local parts of 1PI vertices, then $\Sigma d_{\bar {c}}{\bar
{c}}\partial_{\bar {c}}=\bar{\mu}\partial _{\bar{\mu}}$ induces
the insertion of all the vertex operators $\{O\}$, i.e.,
$\Sigma_{\{O\}}\delta _O\hat{I}_O$ \FL
\begin{eqnarray}
\bar{\mu}\partial _{\bar{\mu}}\Gamma ^{(n)}\left( \left( p\right)
,\left( g\right) ;\{\bar{\mu},[\bar{c}^0] \}\right)
=\Sigma_{\{O\}}\delta _O\hat{I}_O\Gamma ^{(n)}\left( \left(
p\right) ,\left( g\right) ;\{\bar{\mu} ,[\bar{c}^0] \}\right) .
\end{eqnarray}
This is just the general form of renormalization group equation
(RGE) in our approach. Close investigation of the solutions of
Eq.(4) in terms of masses will show that the anomalous dimension
$\delta _O$ of a vertex operator $O$
must be functions of dimensionless tree couplings $[g^0] $ and $%
[\bar{c}^0] $, i.e., $\delta _O=\delta _O\left( [g^0] ,[\bar{c}^0]
\right) $\cite{YY}. The insertion of all the Lagrangian operators
with couplings $\left( g\right) $ can be realized by $g\partial
_g$ (for mass, it is ${m^k}\partial _{m^k},k=1(\text{fermion}
),2(\text{boson})$), i.e., $\Sigma_{\{O\}}\delta
_O\hat{I}_O=\Sigma \delta _gg\partial _g+\Sigma \delta _\phi
\hat{I}_{\partial \phi \partial \phi }+\Sigma_{\{\bar{O}\}}\delta
_{\bar{O}}\hat{I}_{\bar{O}},$ with $\phi $ and $ \bar{O}$ denoting
respectively the 'elementary' fields in Lagrangian and the
operators not defined in Lagrangian. (Here we use $\partial \phi
\partial \phi $ to refer to the kinetic vertex for both fermionic
and bosonic fields of any spin for simplicity, this does not
affect the following deduction as the kinetic terms must be
quadratic in the field operators.) Apparently
$\Sigma_{\{\bar{O}\}}\delta _{ \bar{O}}\hat{I}_{\bar{O}}$ is
absent in renormalizable theories, while for unrenormalizable
models, there will be infinitely many $\bar{O}$ operators. The
insertion of the kinetic operator $\delta _\phi \hat{I}_{\partial
\phi \partial \phi }$ will induce a rescaling of the field
operator $\phi $ by amount $\frac{\delta _\phi }2$. Thus in
renormalizable theories, we obtain that \FL
\begin{equation}
\label{RG}
\left\{ \bar{\mu}\partial _{\bar{\mu}}-\Sigma \bar{\delta}_gg\partial _g-%
\bar{\delta}_{\Gamma ^{\left( n\right) }}\right\} \Gamma
^{(n)}\left( \left( p\right) ,\left( g\right)
;\{\bar{\mu},[\bar{c}^0] \}\right) =0
\end{equation}
with $\bar{\delta}_g\equiv \delta _g-\Sigma _{\left[ \phi \right] _g}\frac{%
\delta _\phi }2$. Since $\left( g\right) $ and $\{\bar{\mu},\left[ \bar{c}%
^0\right] \}$ should be uniquely determined by QTOE, the variation
in Eq.(~\ref{RG}) should be understood as the change due to the
global rescaling of everything. Thus by introducing a natural set
of scale co-moving (or 'running') parameters basing on Coleman's
bacteria analogue\cite{Cole}, we finally arrive at the standard
form of RGE which replaces Eq.(~\ref{RG}) \FL
\begin{equation}
\label{RG2}
\left\{ \mu \partial _\mu -\Sigma \bar{\delta}_{\bar{g}}\bar{g}\partial _{%
\bar{g}}-\bar{\delta}_{\Gamma ^{\left( n\right) }}\right\} \Gamma
^{(n)}\left( \left( p\right) ,\left( \bar{g}\right) ;\{\mu ,[ \bar{c}%
^0] \}\right) =0,\ \ \
\end{equation}
with $\mu \partial _\mu \bar{g}\left( \mu ;\left( g\right) \right) =\bar{g}%
\left( \mu ;\left( g\right) \right) \bar{\delta}_{\bar{g}}\left( \left[ \bar{%
g}^0\left( \mu ;[g^0] \right) \right] ,[\bar{c}^0]
\right) ,$ $\ \bar{g}\left( \mu ;\left( g\right) \right) |_{\mu =\bar{\mu}%
}=g,\ \ \mu \equiv t\bar{\mu},\ \hspace{0.01in}t:\max \left[ \mu
\right] \ll \inf \left\{ \sigma \right\} $. Now we see that the
'running' of the parameters is closely related to the rescaling
procedure of $\bar {\mu}$ whose appearance is naturally guaranteed
in QTOE by the low energy limit operation, the mystery atmosphere
around the dimensional transmutation phenomenon is therefore
removed.

Inserting Eq.(~\ref{RG2}) back into Eq.(~\ref{CSE}) we will get
the full scaling law due to Callan-Symanzik\cite{CS} \FL
\begin{eqnarray}
&&\left\{ s\partial _s+\Sigma
\bar{\delta}_{\bar{g}}\bar{g}\partial_{\bar{g}}+
\bar{\delta}_{\Gamma ^{\left( n\right) }} -d_{\Gamma
^{(n)}}\right\} \Gamma ^{(n)}\left( \left( sp\right) ,\left(
\bar{g}\right) ;\{\bar{\mu},[\bar{c}^0] \}\right) =-i\Gamma
_\Theta ^{(n)}\left( 0,\left( sp\right) ,\left( \bar{g}\right) ;\{
\bar{\mu},[\bar{c}^0] \}\right),\nonumber \\
\end{eqnarray}
where
\FL
\begin{eqnarray}
&&s\partial _s\bar{g}\left( s\bar{\mu};\left( g\right) \right)
 =\bar{g}\left( s\bar{\mu};\left( g\right) \right)
 \bar{\delta}_{\bar{g}}\left(
[ \bar{g}^0( s\bar{\mu};[g^0]) ] ,[\bar{c}^0] \right),\ \ \
\bar{g}\left(s\bar{\mu};( g) \right) |_{s=1}=g,\\ & & i\Gamma
_\Theta ^{(n)}\left( 0,\left( sp\right) ,\left( \bar{g}\right)
;\{\bar{\mu},[\bar{c}^0] \}\right)\equiv \Sigma
d_{\bar{g}}\bar{g}\partial _{\bar{g}}\Gamma ^{(n)}\left( \left(
sp\right) ,\left( \bar{g}\right) ;\{\bar{\mu},[\bar{c}^0]
\}\right) ,
\end{eqnarray}
with $\Theta $ being the trace of the energy tensor of the theory.
Of course in reality we are forced to replace $\{\bar{c%
}\}$ with $\{C\}=\{\mu _{int},\left[ C^0\right] \}$\footnote{Here
$\mu _{int},\left[ C^0\right]$ parallel $\bar{\mu},[\bar{c}^0]$
with $\mu _{int}$ standing for the dimensional constant scale that
will necessarily appear in the indefinite momentum integration and
$\left[ C^0\right]$ for dimensionless constants.}, but in
principle we can start with tree parameters and determine $\{C\}$
by confronting our calculations with experimental data as
mentioned above.

One might oppose that the QTOE is never seen. Our answer is that
the present QFT or field equations has been verified only within a
limited region of the phase space. As a matter of fact, it is no
harm to start with a postulated underlying regularity. If there is
no need of this property, then one can freely remove it at any
time. While in the present QFTs, we do need such underlying
regularity.

Before ending our presentation, we show that RGE can be
interpreted as a decoupling theorem of the underlying high energy
modes. This is easy to see: since the constants $\{\bar {c}\}$
arise from the low energy limit operation, then before taking this
limit Eq.(~\ref{CSE}) must be written as \FL
\begin{eqnarray}
&&\left\{ s\partial _s+\Sigma d_gg\partial _g+\Sigma d_\sigma \sigma
\partial _\sigma -d_{\Gamma ^{(n)}}\right\} \bar{%
\Gamma}^{(n)}(\left( sp\right) ,\left( g\right) ;\{\sigma \}) =0
\end{eqnarray}with the $\Sigma d_{\bar {c}}{\bar
{c}}\partial_{\bar {c}}$ or $\bar{\mu}\partial _{\bar{\mu}}$
replaced by $\Sigma d_\sigma \sigma
\partial _\sigma $. \textbf{This is obviously a normal scaling law in
QTOE.} However, as the constants $\left\{\sigma\right\}$ are
vanishingly small the high energy modes become 'formally'
decoupled while their contributions in the scaling law persist and
appear as 'anomalies' in terms of the 'tree' parameters $\{g\}$,
i.e., the terms $\left\{\Sigma
\bar{\delta}_{\bar{g}}\bar{g}\partial _{
\bar{g}}+\bar{\delta}_{\Gamma ^{\left( n\right) }}\right\} \Gamma
^{(n)}\left( \cdots\right)$ in Eq.(~\ref{RG2}), which are subsumed
into $\bar{\mu}\partial _{\bar{\mu}}$, a coarse grained way to
reproduce the underlying structures' contributions according to
Eq.(~\ref{RG2}),  \FL
\begin{eqnarray}
\label{decoupl}  L_{\left\{ \sigma \right\} }\left(\Sigma d_\sigma
\sigma\partial _\sigma \bar{%
\Gamma}^{(n)}(\cdots)\right)= \left\{\Sigma
\bar{\delta}_{\bar{g}}\bar{g}\partial _{
\bar{g}}+\bar{\delta}_{\Gamma ^{\left( n\right) }}\right\} \Gamma
^{(n)}\left( \cdots\right)= \bar{\mu}\partial _{\bar{\mu}}
\Gamma^{(n)}(\cdots).
\end{eqnarray}Thus it is the
'decoupling effects of high energy modes' that lead to the
violation of naive scaling law in terms of the QFT parameters
$\{g\}$, no divergence is involved here. Since such scaling
'anomalies' can be absorbed into the redefinition of the tree
parameters, we see that such \textbf{'anomalies' from the
underlying modes decoupling lead to finite 'renormalization' of
the tree parameters}. One might understand this mechanism from the
decoupling effects of heavy fermions upon the beta function in QCD
or QED as was illustrated in Ref.\cite{PLB393}, where
$\lim_{M\rightarrow\infty}M\partial_{M}\Gamma=\Delta\beta\alpha
\partial_{\alpha} \Gamma$. Thus
the physical meaning of RGE is deepened in the new strategy,
namely, RGE is an inevitable consequence of the fact that QFTs are
incomplete formulations of the low energy sectors of a complete
theory.
\section{Remarks and summary}
Conventionally, one is forced to use some artificial regulators to
define the UV or high energy ends of a QFT so that the loop
amplitudes could be calculated. Such procedures do not
automatically make the loop amplitudes finite and subsequent
subtraction of divergent pieces is necessary. The subtraction
leads to a residual ambiguity that is in fact encoded in RGE. The
order of logic is first (1) regularization, then (2)
subtraction/renormalization and finally (3) renormalization group.
While in the underlying theory (QTOE) scenario, the amplitudes
that are only formally defined in QFT are coarse grained ones in
QTOE with high energy details integrated out or coarse grained
away. In the decoupling (or low energy) limit, the coarse grained
objects or sectors are subject to certain freedom of redefinition,
which is just the freedom corresponding to renormalization group.
No extra procedures for regularization and subtraction are needed
here except the natural coarse graining and decoupling limit
operation. Had we established the QTOE first in history, RGE would
still be a natural corollary following from coarse graining and
decoupling limit operation, perhaps under a different name. In a
sense, we provide a more physical and reasonable foundation for
renormalization group and finite renormalization without the
somewhat unnatural procedures of regularization and subtraction of
infinities.

We remind that the present formulation are only valid provided the
underlying modes' typical time scale is vanishingly small in
comparison with the QFT processes' time scale. In this sense the
parameters in QFT's are some kind of collective 'coordinates' of
the coarse grained objects. So our approach is in fact pointing
towards a unified framework for quantization, coarse graining,
renormalization and unification of interactions, at least in the
conceptual sense.

Finally, we stress again the there is no need to introduce counter
terms in the differential equation approach, one only needs to fix
the ambiguities, which are especially important for the
electroweak theory with spontaneous symmetry breaking whose
renormalization is rather complicated\cite{EW}. Nonetheless, since
we have heavily relied upon the effective theory versus underlying
complete theory duality, our strategy could be readily applied to
the effective field theory approach, especially to nonperturbative
problems such as nucleon interactions\cite{Nucl}. Further
applications in these perspectives will be pursued in the future.

In summary, we provided natural arguments in favor of a recently
proposed simple strategy for renormalization. The key point is the
existence of {\em a complete quantum theory of everything} which
contains full information of high energy physics that are lacking
in the present the QFT's. From this QTOE scenario, the
Callan-Symanzik equation and RGE can be derived in a natural way
with the RGE recognized as a decoupling theorem of the high energy
modes that underlie the present QFTs or similar quantum theories.
The conceptual foundation for renormalization group is more
reasonable in the QTOE scenario and is not entangled with
divergence at all.

\section*{Acknowledgement}
The author is grateful to W. Zhu for helpful conversations. This
work is supported in part by the National Nature Science
Foundation of China under Grant No. 10075020 and 10205004.

\end{document}